\documentstyle[aps,prl,twocolumn,epsf,floats]{revtex}
\begin{document}
\draft
\preprint{}
\twocolumn[\hsize\textwidth\columnwidth\hsize\csname
@twocolumnfalse\endcsname
\title{
     Momentum Dependence of Resonant Inelastic X-Ray Scattering Spectrum
                           in Insulating Cuprates
}
\author{
                K. Tsutsui, T. Tohyama, and S. Maekawa
}
\address{
                       Institute for Materials Research,
                  Tohoku University, Sendai 980-8577, Japan
}
\date{Received 26 May 1999}
\maketitle
\begin{abstract}
The resonant inelastic x-ray scattering spectrum in insulating cuprates is
examined by using the exact diagonalization technique on small clusters in
the two-dimensional Hubbard model with second and third neighbor hopping terms.
When the incident photon energy is tuned near the Cu $K$ absorption edges,
we find that the features of the unoccupied upper Hubbard band can be extracted from
the spectrum through an anisotropic momentum dependence.
They provide an opportunity for the understanding of the
different behavior of hole- and electron-doped superconductors.
\end{abstract}
\pacs{PACS numbers: 74.25.Jb, 71.10.Fd, 78.70.Ck}

]
\narrowtext

Resonant inelastic x-ray scattering (RIXS) is developing very rapidly into
a powerful technique to investigate elementary excitations in the strongly
correlated electron systems~\cite{Kao,Butorin,Hill,Kuiper,Abbamonte}.
The application of this technique to insulating copper oxides has made it
possible to observe an excitation due to a local charge
transfer between copper and oxygen~\cite{Hill} and local $d$-$d$ excitations
on copper site~\cite{Kuiper}.
In addition, it has been demonstrated that, by using high resolution
experiments~\cite{Abbamonte}, the momentum-dependent measurement of the charge
transfer gap is possible when the incident photon energy $\omega_i$ is
tuned through Cu $K$ absorption edge.
Thus, the RIXS can be a useful probe to obtain information on the momentum
dependence of the elementary excitations.

One of the elementary excitations in the insulating cuprates is the
charge-transfer process from the occupied Zhang-Rice singlet band
(ZRB)~\cite{Zhang} composed of Cu 3$d_{x^2-y^2}$ and O 2$p_\sigma$ orbitals to
the unoccupied upper Hubbard band (UHB).
The dispersion of ZRB have been extensively studied by angle-resolved
photoemission spectroscopy (ARPES) experiments on the parent compounds of
high $T_c$ superconductors~\cite{Wells,Kim,Ronning}:  A $d$-wave-like
dispersion was observed along the (0,$\pi$)-($\pi$,0) line with the minimum
of the binding energy at ($\pi$/2,$\pi$/2)~\cite{Ronning}.
On the contrary, the dispersion relation and spectral properties of the
unoccupied UHB have not been examined and thus remain to be understood.
The information of UHB is of crucial importance for the understanding of
the motion of electrons in the electron-doped superconductors.
In addition, it may be useful to know if the particle-hole symmetry is
required for the high temperature superconductivity.

In this Letter, we examine the RIXS spectrum for the Cu $K$-edge, and
demonstrate that the characteristic features of the dispersion of UHB can be
extracted from the momentum dependence of the spectrum.
To see this, we use the half-filled single-band Hubbard model to describe
the occupied ZRB and unoccupied UHB by mapping ZRB onto the lower Hubbard band
(LHB) in the model.
Then, we incorporate Cu 1$s$ and 4$p$ orbitals into the model to include the
1$s$-core hole and excited 4$p$ electron into the intermediate state of the
RIXS process.
The long-range hoppings are also introduced in the Hubbard model with realistic values
obtained from the analysis of ARPES data.
We find a characteristic momentum dependence of the Cu $K$-edge RIXS spectrum:
The energy of the threshold of the RIXS spectrum at ($\pi$/2,$\pi$/2) is higher
than that at (0,0), whereas the energy of the threshold at ($\pi$/2,0)
is lower than that at (0,0).
This anisotropic dependence is explained by the dispersion of the UHB which has
the minimum energy at ($\pi$,0) due to the long-range hoppings.
The determination of the UHB will contribute to the understanding of the
different behavior of hole- and electron-doped superconductors~\cite{Kim,Takagi}.

We map the ZRB onto the LHB, which is equivalent to the
elimination of O $2p$ orbitals.
Such mapping was used in the analysis of O $1s$ x-ray absorption
spectrum\cite{Chen}.
The Hubbard Hamiltonian with second and third neighbor hoppings for the $3d$
electron system is written as,
\begin{eqnarray}\label{ham3d}
H_{3d} &=& -t\sum_{\langle {\bf i},{\bf j} \rangle_{\rm 1st}, \sigma}
          d_{{\bf i},\sigma}^\dag d_{{\bf j},\sigma}
        -t'\sum_{\langle {\bf i},{\bf j} \rangle_{\rm 2nd}, \sigma}
          d_{{\bf i},\sigma}^\dag d_{{\bf j},\sigma} \nonumber\\
      &&-t''\sum_{\langle {\bf i},{\bf j} \rangle_{\rm 3rd}, \sigma}
          d_{{\bf i},\sigma}^\dag d_{{\bf j},\sigma} + {\rm H.c.}
      +U\sum_{\bf i}
          n^d_{{\bf i},\uparrow}n^d_{{\bf i},\downarrow},
\end{eqnarray}
where $d_{{\bf i},\sigma}^\dag$ is the creation operator of $3d$ electron
with spin $\sigma$ at site ${\bf i}$,
$n_{{\bf i},\sigma}^d=d_{{\bf i},\sigma}^\dag
d_{{\bf j},\sigma}$, the summations
$\langle {\bf i},{\bf j} \rangle_{\rm 1st}$,
$\langle {\bf i},{\bf j} \rangle_{\rm 2nd}$, and
$\langle {\bf i},{\bf j} \rangle_{\rm 3rd}$ run over first, second,
and third nearest-neighbor pairs, respectively, and the rest of the notation
is standard.

\begin{figure}
\epsfxsize=7cm
\centerline{\epsffile{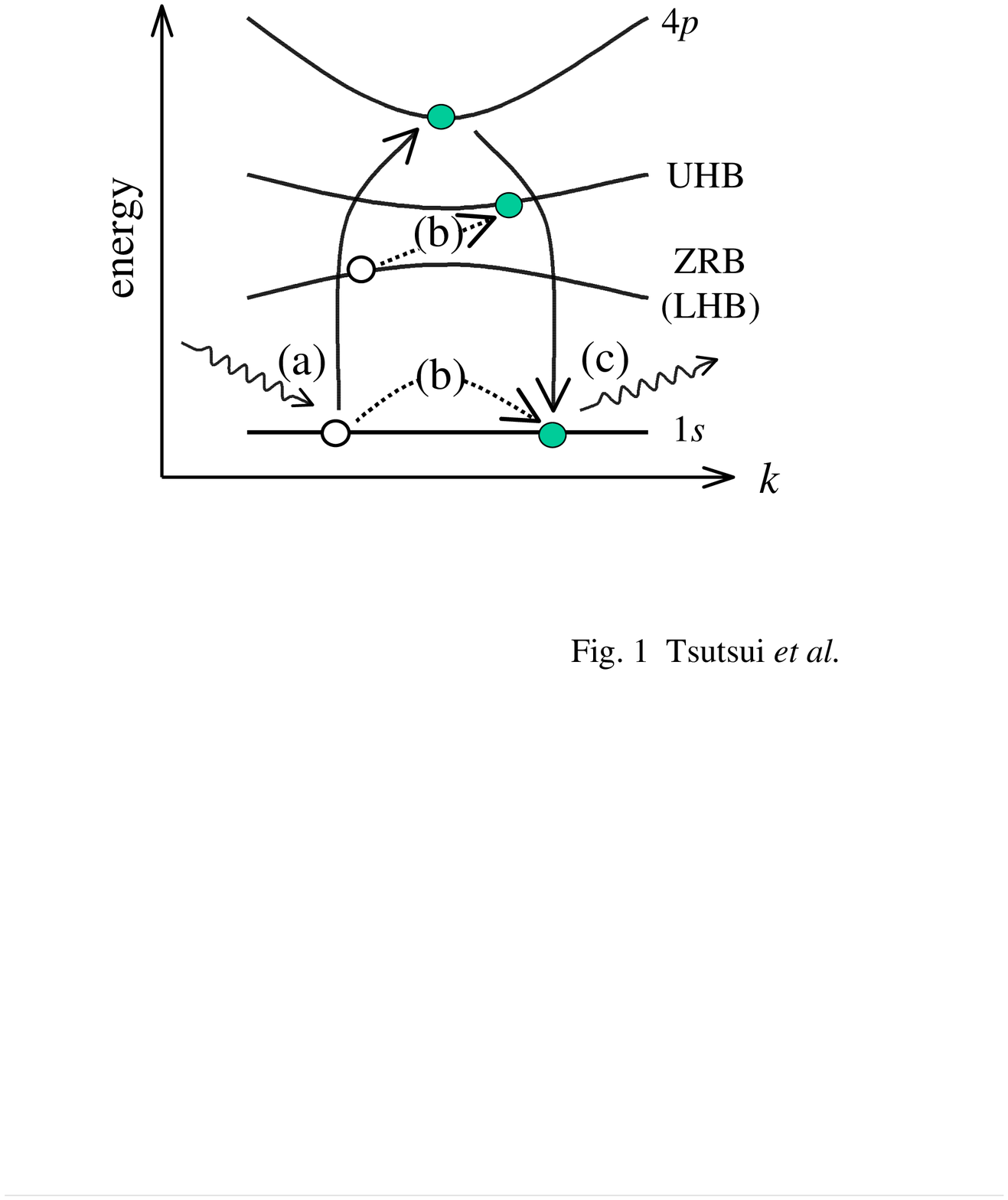}}
\caption{Schematic picture of the Cu K-edge RIXS process.
An incident photon is absorbed and dipole transition $1s\rightarrow 4p$ is
 brought about [process (a)], and through the intermediate state
[process (b)], the photo-excited $4p$ electron goes to $1s$ again and
a photon is emitted [process (c)].}
\label{figpic}\end{figure}

Figure~\ref{figpic} shows the schematic process of Cu $K$-edge RIXS.
An absorption of an incident photon with energy $\omega_i$, momentum
${\bf K}_i$, and polarization ${\bf \epsilon}_i$ brings about
the dipole transition of an electron from Cu $1s$ to $4p$ orbital
[process (a) in Fig.~\ref{figpic}].
In the intermediate states, $3d$ electrons interact with a $1s$-core hole and
a photo-excited $4p$ electron via the Coulomb interactions so that the
excitations in the $3d$ electron system are evolved
[process (b)].
The $4p$ electron goes back to the $1s$ orbital again and a photon with
energy $\omega_f$, momentum ${\bf K}_f$, and polarization
${\bf\epsilon}_f$ is emitted [process (c)].
The differences of the energies and the momenta between incident and emitted
photons are transferred to the $3d$ electron system.

%%%%%%%%%%%%%%%%
%%%%%%%%%%%%%%%%%%%%%%%%%%%%%%%%%%%%%%%%%%%%%%%%%%%%%%%%%%%%%%%%%%%%%%%%%%%%%%%%%%

In the intermediate state, there are a $1s$-core hole and a $4p$ electron,
with which $3d$ electrons interact.
Since the 1$s$-core hole is localized because of a small radius of the Cu
1$s$ orbital, the attractive interaction between the 1$s$-core hole and
3$d$ electrons is very strong.
The interaction is written as,
\begin{eqnarray}
H_{1s\text{-}3d}=-V\sum_{{\bf i},\sigma,\sigma'}
n_{{\bf i},\sigma}^d n_{{\bf i},\sigma'}^s,
\end{eqnarray}
where $n_{{\bf i},\sigma}^s$ is the number operator of 1$s$-core hole
with spin $\sigma$ at site ${\bf i}$, and $V$ is taken to be positive.
On the contrary, since the 4$p$ electron is delocalized, the repulsive
interaction between the 4$p$ and 3$d$ electrons as well as the attractive
one between the 4$p$ electron and the 1$s$-core hole is small as compared
with the 1$s$-3$d$ interaction.
In addition, when the core-hole is screened by the 3$d$ electrons through
the strong 1$s$-3$d$ interaction,
effective charge that acts on the 4$p$ electron at the core-hole site
becomes small.
Therefore, the interactions related to the 4$p$ electron are neglected for simplicity.
Furthermore, we assume that the photo-excited 4$p$ electron enters into
the bottom of the 4$p_z$ band with momentum ${\bf k}_0$, where $z$-axis is 
perpendicular to the CuO$_2$ plane.
This assumption is justified as long as the Coulomb interactions associated
with the 4$p$ electron are neglected and the resonance condition is set to
the threshold of the 1$s$$\rightarrow$4$p_z$ absorption spectrum\cite{polarization}.
Under these assumptions, the RIXS spectrum is expressed as,
\begin{eqnarray}\label{rixs}
I(\Delta {\bf K},\Delta\omega)&=&\sum_\alpha\left|\langle\alpha|
\sum_\sigma s_{{\bf k}_0-{\bf K}_f,\sigma} p_{{\bf k}_0,\sigma}
\right.\nonumber\\&&\times\left.
\frac{1}{H-E_0-\omega_i-i\Gamma}
p_{{\bf k}_0,\sigma}^\dag s_{{\bf k}_0-{\bf K}_i,\sigma}^\dag
|0\rangle\right|^2
\nonumber\\&&\times
\delta(\Delta\omega-E_\alpha+E_0),
\end{eqnarray}
where $H=H_{3d}+H_{1s\text{-}3d}+H_{1s,4p}$, $H_{1s,4p}$ being
kinetic and on-site energy terms  for a 1$s$-core hole and a 4$p$ electron,
$\Delta{\bf K}={\bf K}_i-{\bf K}_f$,
$\Delta\omega=\omega_i-\omega_f$,
$s_{{\bf k},\sigma}^\dag$ ($p_{{\bf k},\sigma}^\dag$) is the creation operator
of the 1$s$-core hole (4$p$ electron) with momentum ${\bf k}$ and spin $\sigma$,
$|0\rangle$ is the ground state of the half-filled system with energy $E_0$,
$|\alpha\rangle$ is the final state of the RIXS process with energy $E_\alpha$, and
$\Gamma$ is the inverse of the relaxation time in the intermediate state.
In Eq.~(\ref{rixs}), the terms $H_{1s,4p}$ are replaced by $\varepsilon_{1s\text{-}4p}$
which is the energy difference between the $1s$ level and the bottom of the $4p_z$ band.

The RIXS spectrum of Eq.~(\ref{rixs}) is calculated on $(\sqrt 8$$\times$$\sqrt 8)$-,
$(\sqrt{10}$$\times$$\sqrt{10})$-, and (4$\times$4)-site
clusters with periodic boundary condition by using a modified version of
the conjugate-gradient method together with the Lancz\"os technique.
We will show the results for the 4$\times$4-site cluster in the following.

The values of the parameters are as follows:
$t'/t=-0.34$, $t''/t=0.23$, $U/t=10$, $V/t=15$,
and $\Gamma/t=1$ with $t=0.35$~eV.
The values of $t$, $t'$, and $t''$ are the same as those used in the
analysis of ARPES data of the high $T_c$ superconductors
based on the $t$-$t'$-$t''$-$J$ model~\cite{Kim,Tiny}.
The value of $U$ is obtained from the relation $J$=4$t^2$/$U$ and $J$/$t$=0.4.
The value of $V$ is set to be larger than that of $U$.
The results shown below are insensitive to the magnitude of $V$ as well as
of $\Gamma$~\cite{plat98}.

\begin{figure}
\epsfxsize=7cm
\centerline{\epsffile{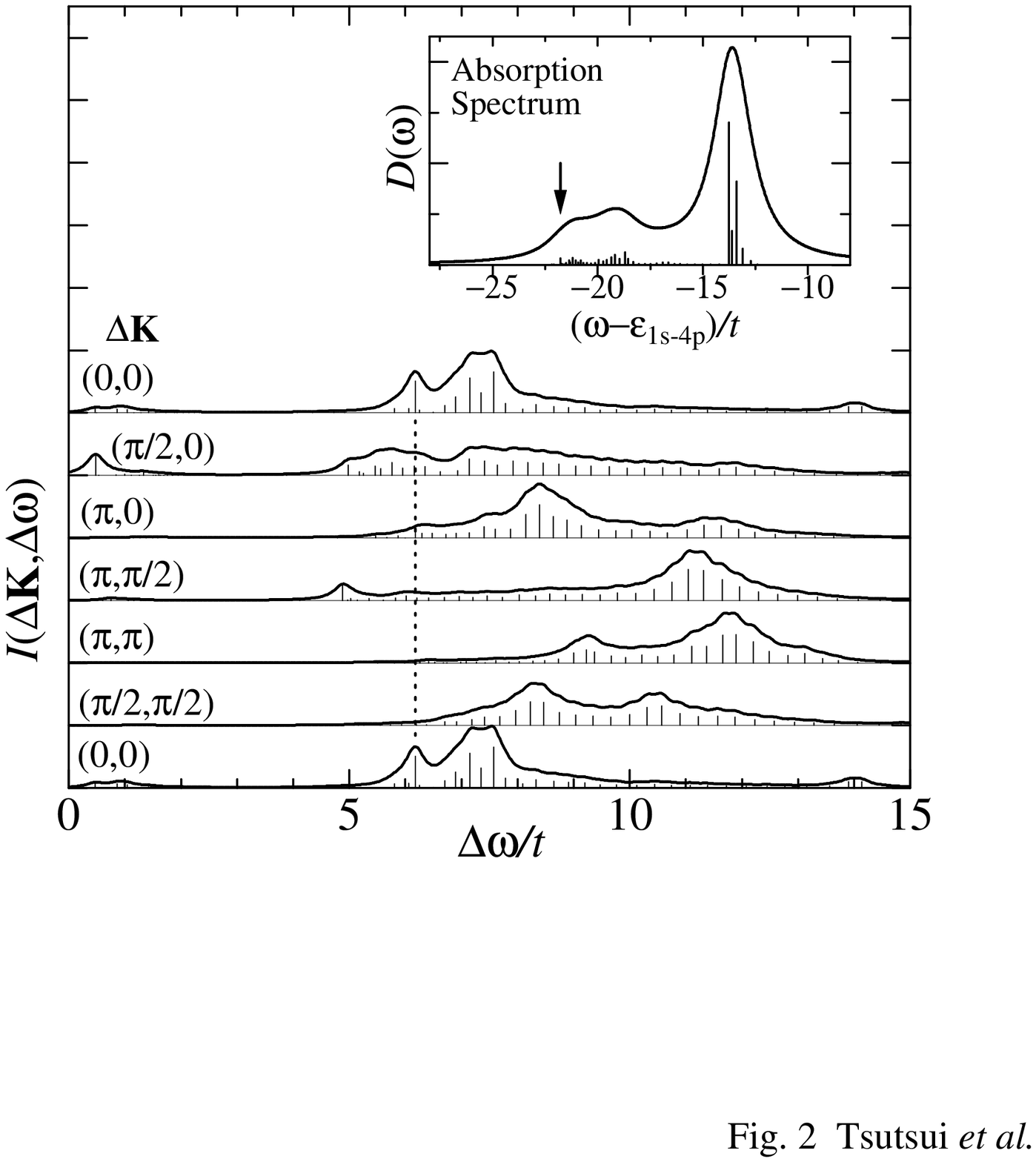}}
\caption{
Resonant inelastic x-ray scattering spectra for Cu $K$-edge
in half-filled Hubbard model with long-range hoppings.  The spectra of the elastic
scattering process at $\Delta{\bf K}$=(0,0) are not shown.
The parameters used are $U/t=10$, $V/t=15$, $\Gamma/t=1$,
$t'/t=-0.34$, and $t''/t=0.23$.
The vertical dotted line denotes the position of the peak at
$\Delta{\bf K}=(0,0)$ for guide to eyes.
The $\delta$-functions (the vertical thin solid lines) are convoluted
with a Lorentzian broadening of 0.2$t$.
Inset is the Cu 1$s$ absorption spectrum with $\Gamma_{\text{XAS}}/t=\Gamma/t=1.0$,
and the incident photon energy $\omega_i$ is set to the value denoted by the arrow.}
\label{figrixs}\end{figure}

Before going into the RIXS spectrum, we mention the resonance condition
in RIXS.
To determine the condition, we have to examine the Cu 1$s$
x-ray absorption spectroscopy (XAS) spectrum defined as,
\begin{eqnarray}
D(\omega)&=&\frac{1}{\pi}{\rm Im}\langle 0 |
s_{{\bf k}_0-{\bf K}_i,\sigma} p_{{\bf k}_0,\sigma}
\frac{1}{H-E_0-\omega-i\Gamma_{\text{XAS}}}
\nonumber\\&&\times
p_{{\bf k}_0,\sigma}^\dag s_{{\bf k}_0-{\bf K}_i,\sigma}^\dag
|0\rangle,
\end{eqnarray}
where $H$ is the same as that in Eq.~(\ref{rixs}).
It is necessary to tune the incident photon
energy $\omega_i$ to the energy region where the Cu 1$s$ XAS spectrum appears.
The inset in Fig.~\ref{figrixs} shows $D(\omega)$, where a two-peak structure appears,
{\it i.e.}, the broad one around $\omega-\varepsilon_{1s\text{-}4p}=-20t$
and the sharp one around $-13t$.
The former mainly contains configurations that the core-hole site is doubly
occupied by the 3$d$ electrons ($U-2V=-20t$),
while the latter dominantly contains configurations that the core-hole site
is singly occupied ($-V=-15t$).
This means that, when the incident energy is tuned around the former
structure, the information about UHB can be extracted from the RIXS spectrum.
Thus, we set $\omega_i$ to the threshold of the XAS spectrum denoted by
the arrow in the inset.

Figure~\ref{figrixs} shows the momentum dependence of the RIXS spectrum.
The spectra below $\Delta\omega/t\sim2$ and above $\Delta\omega/t\sim5$
have different origins:  The former comes from the excitations related to the
spin degree of freedom such as two-magnon Raman scattering, the energy scale
of which is so small that the spectrum is hard to be observed\cite{Abbamonte}.
On the other hand, the latter is related to the excitations from LHB to UHB.
The vertical dotted line in the figure denotes the position of the low-energy
peak at $\Delta{\bf K}=(0,0)$ for guide to eyes.
The spectra strongly depend on the momentum showing a feature that the
weight shifts to higher energy region with increasing
$\left|\Delta{\bf K}\right|$.
This momentum dependence is also obtained in the 10-site cluster calculations.
In addition, the threshold of the spectrum at
$\Delta{\bf K}=(\pi/2,0)$ and ($\pi$,$\pi$/2) is lower in energy
than that at (0,0).
At $\Delta{\bf K}=(\pi/2,\pi/2)$, however, the spectrum appears above
the threshold at (0,0), resulting in an anisotropic momentum dependence
between the spectra along (0,0) to ($\pi$/2,0) and along (0,0) to
($\pi$/2,$\pi$/2).
We note that such an anisotropic feature can not be observed in the RIXS spectrum
of the Hubbard model without $t'$ and $t''$ (not shown)\cite{Typical}.
In the 8-site cluster the spectrum at $\Delta{\bf K}=(\pi/2,\pi/2)$
is lower in energy than that at ($\pi$,0), which is consistent with the above feature.

%%%%%%%%%%%%%%%%
%%%%%%%%%%%%%%%%%%%%%%%%%%%%%%%%%%%%%%%%%%%%%%%%%%%%%%%%%%%%%%%%%%%%%%%%%%%%%%%%%%

In the intermediate state, the excitations of the 3$d$ electrons from the occupied
to unoccupied states are caused by the interaction with the 1$s$-core hole,
{\it i.e.}, $H_{1s\text{-}3d}=-V/N\sum_{{\bf k}_1\sim{\bf k}_4,\sigma,\sigma'}
\delta_{{\bf k}_2-{\bf k}_1,{\bf k}_3-{\bf k}_4}
d_{{\bf k}_1,\sigma}^\dag d_{{\bf k}_2,\sigma}
s_{{\bf k}_3,\sigma'}^\dag s_{{\bf k}_4,\sigma'}$.
The operator $d_{{\bf k}_1,\sigma}^\dag d_{{\bf k}_2,\sigma}$
represents a particle-hole excitation.
Therefore, we analyze the RIXS process by decomposing into such a
particle-hole excitation in order to understand the anisotropic momentum dependence
of the threshold along (0,0) to ($\pi$/2,0) and along (0,0) to ($\pi$/2,$\pi$/2)
in Fig.~\ref{figrixs}.

\begin{figure}
\epsfxsize=7cm
\centerline{\epsffile{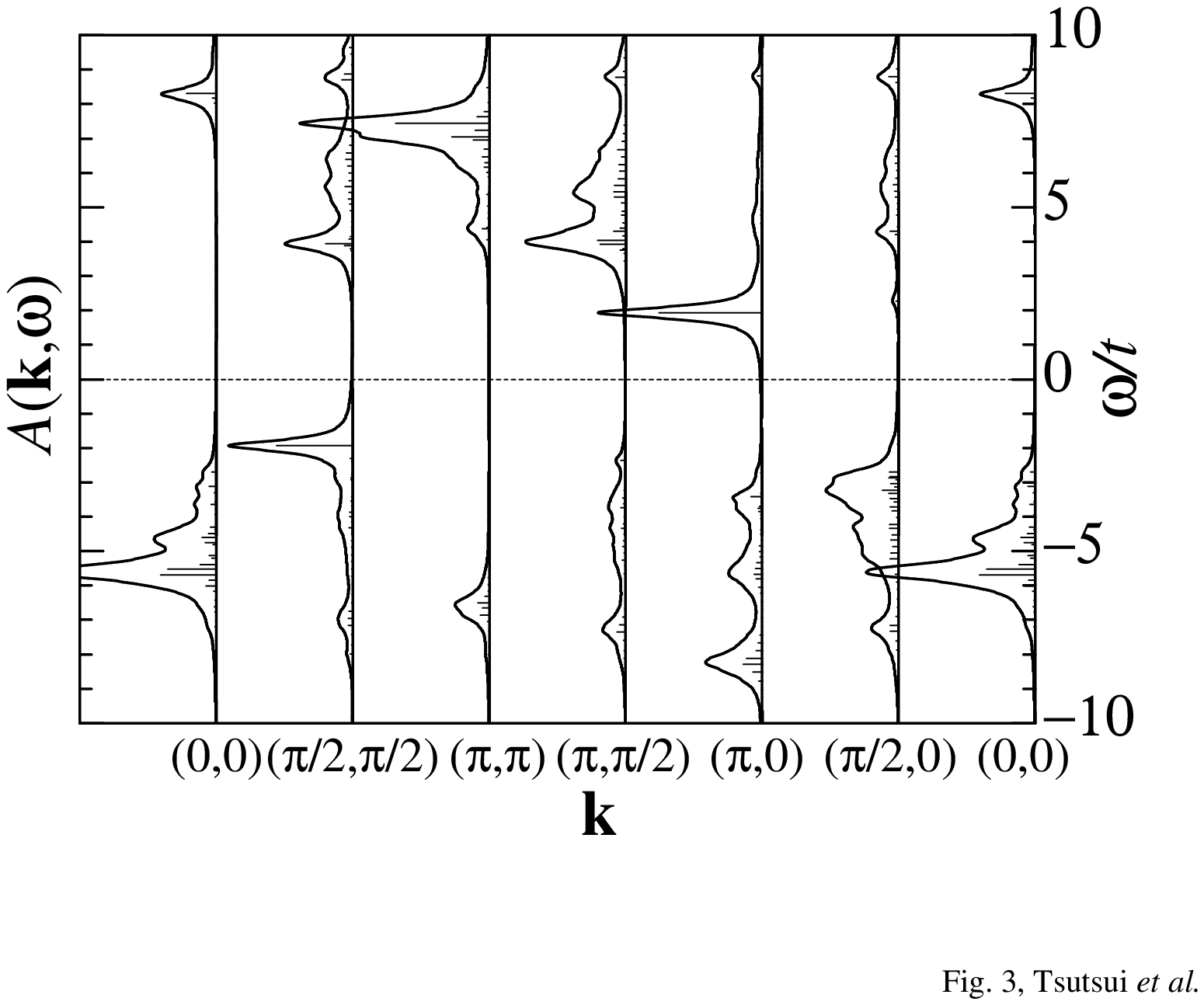}}
\caption{Single particle excitation spectrum $A({\bf k},\omega)$ in the half-filled
Hubbard model with $t'$ and $t''$ terms.
$t'/t=-0.34$ and $t''/t=0.23$.
The dotted line denotes the chemical potential.
The $\delta$ functions are convoluted with a Lorentzian broadening of 0.2$t$.}
\label{figakw}\end{figure}

As a first step, we consider the particle-hole excitation as the convolution of the
single-particle excitation spectra $A({\bf k},\omega)$ between occupied LHB and
unoccupied UHB.
Figure~\ref{figakw} shows $A({\bf k},\omega)$ in the half-filled
Hubbard model with $t'$ and $t''$ terms\cite{hubakw}.
Below the chemical potential denoted by the dotted line, a sharp peak appears
at ($\pi$/2,$\pi/$2) with the lowest-binding energy.
In contrast, the spectrum at ($\pi$,0) is very broad and deep in energy.
These features are consistent with the ARPES data for Sr$_2$CuO$_2$Cl$_2$\cite{Wells,Kim}.
Above the chemical potential, the dispersion of UHB has the minimum of the
energy at ${\bf k}$=($\pi$,0)\cite{SCB}.
We show below that the ($\pi$,0) spectrum in UHB plays a crucial role in the
RIXS spectrum with $\Delta{\bf K}$=($\pi$/2,0).

%%%%%%%%%%%%%%%%
%%%%%%%%%%%%%%%%%%%%%%%%%%%%%%%%%%%%%%%%%%%%%%%%%%%%%%%%%%%%%%%%%%%%%%%%%%%%%%%%%

\begin{figure}
\epsfxsize=7cm
\centerline{\epsffile{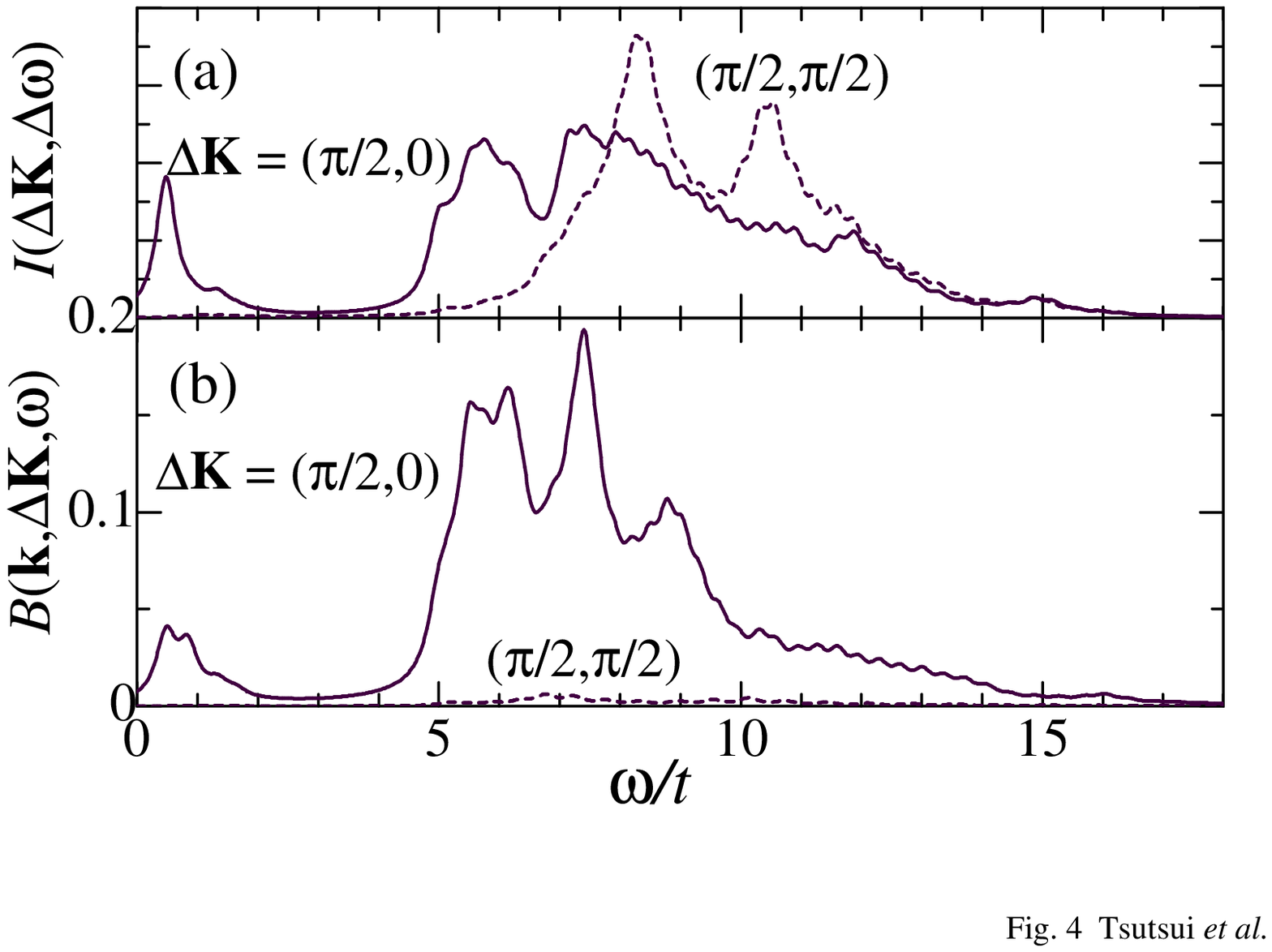}}
\caption{Spectral function $B({\bf k},\Delta{\bf K},\omega)$
which represents the particle-hole excitation process (lower panel),
with $[{\bf k},\Delta{\bf K}]=[(\pi/2,0),(\pi/2,0)]$
(solid line) and $[(\pi/2,-\pi/2),(\pi/2,\pi/2)]$ (dotted line).
Upper panel shows the corresponding RIXS spectra $I(\Delta{\bf K},\Delta\omega)$ which
are given in Fig.~\ref{figrixs}.}
\label{figtwo}\end{figure}

Now, we examine the lowest-energy excitations with
$\Delta{\bf K}$=(0,0), ($\pi$/2,0), and ($\pi/$2,$\pi$/2) in the convoluted spectrum
$\int_{{\cal E}\le\mu} d{\cal E}A({\bf k}+\Delta{\bf K},{\cal E}+\omega)
A({\bf k},{\cal E})$, where $\mu$ is the chemical potential.
For the case that $\Delta{\bf K}$=(0,0), the minimum excitation energy
in the convoluted spectrum is $\sim$5$t$ at ${\bf k}$=($\pi$,0).
In the same way, the lowest-energy excitation with $\Delta{\bf K}$=($\pi$/2,0) in
the convoluted spectrum is that from ${\bf k}$=($\pi$/2,0) of LHB to
($\pi$,0) of UHB with the energy of $\sim$4$t$.
This value is smaller than that for the $\Delta{\bf K}$=(0,0) case,
being consistent with the relation of the thresholds of the RIXS
spectra between $\Delta{\bf K}$=(0,0) and ($\pi$/2,0).
In contrast, the lowest-energy excitation with $\Delta{\bf K}$=($\pi$/2,$\pi$/2)
in the convoluted spectrum is inconsistent with that in the RIXS spectrum,
because the excitation energy, which is determined by the peaks at
${\bf k}$=($\pi$/2,$-\pi$/2) of LHB and at ($\pi$,0) of UHB,
is almost the same as that for $\Delta{\bf K}$=($\pi$/2,0).
This means that the argument based on the convoluted spectrum of $A({\bf k},\omega)$
is insufficient to understand the anisotropic behavior of the threshold,
and thus we need to treat the process of the particle-hole excitation exactly.
Therefore, we introduce the spectral function $B({\bf k},\Delta{\bf K};\omega)$
of the two-body Green's function, which can describe the
particle-hole excitation, defined as
\begin{eqnarray}
B({\bf k},\Delta{\bf K};\omega)&=&
\sum_\alpha\left|\langle\alpha|\sum_\sigma
d_{{\bf k}+\Delta{\bf K},\sigma}^\dag d_{{\bf k},\sigma}|0\rangle\right|^2
\nonumber\\&&\times
\delta(\omega-E_\alpha+E_0),
\end{eqnarray}
where the states $|\alpha\rangle$ have the same point-group symmetry
as that of the final states of the RIXS process.
Figure~\ref{figtwo} shows $B({\bf k},\Delta{\bf K};\omega)$ with
$[{\bf k},\Delta{\bf K}]=[(\pi/2,0),(\pi/2,0)]$ and
$[(\pi/2,-\pi/2),(\pi/2,\pi/2)]$.
Note that ${\bf k}+\Delta{\bf K}=(\pi,0)$ in both cases.
The spectrum with $[{\bf k}, \Delta{\bf K}]=[(\pi/2,0),(\pi/2,0)]$
reproduces very well the RIXS spectrum near the threshold.
The intensity of $B({\bf k},\Delta{\bf K};\omega)$ with
$\Delta{\bf K}=(\pi/2,\pi/2)$ is very small, implying that
the process $\sum_\sigma d_{(\pi,0),\sigma}^\dag d_{(\pi/2,-\pi/2),\sigma}$
is almost forbidden\cite{SDW}.
This is the origin of the small weight of $I(\Delta{\bf K},\omega)$ with
$\Delta{\bf K}=(\pi/2,\pi/2)$ around $\omega/t\sim 5$.
We note that this behavior can not be obtained by the convolution
of $A({\bf k},\omega)$ mentioned above.
At the higher energy region, $\omega/t\sim 7$, different processes keeping
$\Delta{\bf K}=(\pi/2,\pi/2)$, for example, the annihilation of
the $(\pi/2,0)$ electron and the creation of the $(\pi,\pi/2)$ one,
dominate the excitations.
These processes induce the large weight above $\omega/t\sim 7$.

%%%%%%%%%%%%%%%%
%%%%%%%%%%%%%%%%

In summary, we have examined the momentum dependence of the Cu $K$-edge
RIXS spectrum by using numerically exact diagonalization technique on small
clusters.
Regarding the ZRB as the LHB, we have adopted the Hubbard model with
Cu $1s$- and $4p$-bands.
We have also introduced the long-range hoppings, $t'$ and $t''$, of
the ZR singlet, and found that the threshold of the spectrum at
$\Delta{\bf K}=(\pi/2,0)$ is small
compared with that at (0,0), whereas the threshold at $(\pi/2,\pi/2)$ is
larger than that at (0,0).
This dependence is caused by the ($\pi$,0) state in UHB.
Thus, by examining the RIXS spectrum, we can extract the property of the
unoccupied states that is crucially important for the electron-doped
superconductors and also for the understanding of the behavior different
from the hole-doped superconductors~\cite{Kim,Takagi}.
Very recently, Stanford's group has reported \cite{Hasan} that interesting
data with momentum dependent inelastic scattering signal has been seen in
the 2 eV range, which reveals the property of unoccupied states right above
the charge-transfer gap.
These progress in both theory and experiment will open a new prospect of
the physics of cuprates.

The authors thank Z.-X. Shen for valuable discussions.
This work was supported by Priority-Areas Grants from the Ministry of
Education, Science, Culture and Sport of Japan, CREST, and NEDO.
Computations were carried out in ISSP, Univ. of Tokyo; IMR,
Tohoku Univ.; and Tohoku Univ..

%%%%%%%%%%%%%%%%
%Figure Captions
%%%%%%%%%%%%%%%%

%%%%%%%%%%%
%References
%%%%%%%%%%%


\vspace{-.4cm}
\begin{references}
\vspace{-1.6cm}
%%%%%%%%%%%%%%%%%%%%%%%%%%%%%%%%%%
\bibitem{Kao} C.-C. Kao {\it et al.}, Phys. Rev. B {\bf 54}, 16361 (1996).
%%%%%%%%%%%%%%%%%%%%%%%%%%%%%%%%%%
\bibitem{Butorin} S. M. Butorin {\it et al.}, Phys. Rev. B {\bf 55}, 4242 (1997).
%%%%%%%%%%%%%%%%%%%%%%%%%%%%%%%%%%
\bibitem{Hill} J. P. Hill {\it et al.}, Phys. Rev. Lett. {\bf 80}, 4967 (1998).
%%%%%%%%%%%%%%%%%%%%%%%%%%%%%%%%%%
\bibitem{Kuiper} P. Kuiper {\it et al.}, Phys. Rev. Lett. {\bf 80}, 5204 (1998).
%%%%%%%%%%%%%%%%%%%%%%%%%%%%%%%%%%
\bibitem{Abbamonte} P. Abbamonte {\it et al.}, cond-mat/9810095.
The best resolution used was 0.45 eV.
%%%%%%%%%%%%%%%%%%%%%%%%%%%%%%%%%%
\bibitem{Zhang}  F. C. Zhang and T. M. Rice, Phys. Rev. B {\bf 37}, 3759 (1988).
%%%%%%%%%%%%%%%%%%%%%%%%%%%%%%%%%%
\bibitem{Wells}  B. O. Wells {\it et al.}, Phys. Rev. Lett. {\bf 74}, 964 (1995).
%%%%%%%%%%%%%%%%%%%%%%%%%%%%%%%%%%
\bibitem{Kim}    C. Kim {\it et al.}, Phys. Rev. Lett. {\bf 80}, 4245 (1998).
%%%%%%%%%%%%%%%%%%%%%%%%%%%%%%%%%%
\bibitem{Ronning}  F. Ronning {\it et al.}, Science, {\bf 282}, 2067 (1998).
%%%%%%%%%%%%%%%%%%%%%%%%%%%%%%%%%%
\bibitem{Takagi} H. Takagi {\it et al.}, Physica C {\bf 162-164}, 1001 (1989).
%%%%%%%%%%%%%%%%%%%%%%%%%%%%%%%%%%
\bibitem{Chen} C. T. Chen {\it et al.}, Phys. Rev. Lett. {\bf 66}, 104 (1991).
%%%%%%%%%%%%%%%%%%%%%%%%%%%%%%%%%%
\bibitem{polarization}
The experimental data for the Cu $K$-edge absorption spectrum
show that the out-of-plane ($z$) polarized peaks are lower in energy than
in-plane polarized peaks\cite{Kosugi}.
In order to examine the RIXS spectrum with the single component of the 4$p$ bands,
we consider the case of $z$-polarized photon.
%%%%%%%%%%%%%%%%%%%%%%%%%%%%%%%%%%
\bibitem{Tiny}
In 8- and 10-site clusters, only $t$ and $t'$ are included for hopping parameters due to their small system sizes.
%%%%%%%%%%%%%%%%%%%%%%%%%%%%%%%%%%
\bibitem{plat98}   P. M. Platzman and E. D. Isaacs, Phys. Rev. B {\bf 57}, 11107 (1998).
%%%%%%%%%%%%%%%%%%%%%%%%%%%%%%%%%%
\bibitem{Typical} K. Tsutsui, T. Tohyama, and S. Maekawa (unpublished).
%%%%%%%%%%%%%%%%%%%%%%%%%%%%%%%%%%
\bibitem{hubakw}
For $A({\bf k},\omega)$ in the Hubbard model without $t'$ and $t''$,
see P. W. Leung {\it et al.}, Phys. Rev. B {\bf 46}, 11779 (1992).
%%%%%%%%%%%%%%%%%%%%%%%%%%%%%%%%%%
\bibitem{SCB}
In the large $U$ limit, the dispersion of the UHB is described by the
single-hole dispersion of the $t$-$t'$-$t''$-$J$ model with $t$$<$0,
$t'$$>$0, and $t''$$<$0.  The sign difference comes from the fact that the
carrier in UHB is an electron [see T. Tohyama. and S. Maekawa. Phys. Rev. B
{\bf 49}, 3596 (1994)].  Although the energy at ($\pi$,0) is almost the same
as that at ($\pi$/2,$\pi$/2) in the $t$-$J$ model, the former decreases when
$t'$($>$0) and $t''$($<$0) are introduced.
By defining $\epsilon$(${\bf k}$)=4$t'$cos$k_x$cos$k_y$+2$t''$(cos$k_x$+cos$k_y$),
the energy difference between at ($\pi$/2,$\pi$/2) and ($\pi$,0) is proportional to
$\epsilon$($\pi$/2,$\pi$/2)$-$$\epsilon$($\pi$,0)=$-$8$t''$+4$t'$.
%%%%%%%%%%%%%%%%%%%%%%%%%%%%%%%%%%
\bibitem{SDW}
In the spin-density wave mean-field approximation,
$B({\bf k},\Delta{\bf K};\omega)$ with [${\bf k}$,
$\Delta{\bf K}$]= [($\pi$/2,$-\pi$/2),($\pi$/2,$\pi$/2)] is rigorously zero.
This is due to the effect of the coherence factor arising from the antiferromagnetic
long-range order as is the case of the BCS superconductivity.
The detail will be shown elsewhere.
%%%%%%%%%%%%%%%%%%%%%%%%%%%%%%%%%%
\bibitem{Hasan} Z. Hasan, E. Issacs, and Z.-X. Shen (unpublished).
%%%%%%%%%%%%%%%%%%%%%%%%%%%%%%%%%%
\bibitem{Kosugi} N. Kosugi {\it et al.}, Chem. Phys. {\bf 135}, 149 (1989)
%%%%%%%%%%%%%%%%%%%%%%%%%%%%%%%%%%
\end{references}
\end{document}